\documentclass[
  aip,amsmath,amssymb,
  reprint,
]{revtex4-1}

\usepackage{graphicx}
\usepackage{dcolumn}
\usepackage[T1]{fontenc}
\usepackage[utf8]{inputenc}
\usepackage[colorlinks,breaklinks=true]{hyperref}
\usepackage{mathptmx}
\usepackage{etoolbox}
\usepackage{bm}

\usepackage{multirow}

\hypersetup{
   colorlinks = true,
   linkcolor  = blue,
   citecolor  = blue,
   urlcolor   = blue
}

\newcommand {\crg}{\color{black}}  

\makeatletter
\def\@email#1#2#3{%
 \endgroup
 \patchcmd{\titleblock@produce}
  {\frontmatter@RRAPformat}
  {\frontmatter@RRAPformat{\produce@RRAP{*#1\href{mailto:#2}{#2} and \href{mailto:#3}{#3}}}\frontmatter@RRAPformat}
  {}{}
}
\begin{document}

\preprint{AIP/123-QED}

\title[]{Characterization of broadband Purcell filters with compact footprint for fast multiplexed superconducting qubit readout}

\author{Seong Hyeon Park}
  \affiliation{Department of Electrical and Computer Engineering, Seoul National University, Seoul 08826, Republic of Korea}

\author{Gahyun Choi*}
  \affiliation{Center for Superconducting Quantum Computing Systems, Korea Research Institute of Standards and Science, Daejeon 34113, Republic of Korea}
  \thanks{}

\author{Gyunghun Kim}
  \affiliation{Department of Electrical and Computer Engineering, Seoul National University, Seoul 08826, Republic of Korea}

\author{Jaehyeong Jo}
  \affiliation{Department of Physics, Ulsan National Institute of Science and Technology (UNIST), Ulsan 44919, Republic of Korea}

\author{Bumsung Lee}
  \affiliation{Samsung Electronics, Hwaseong-si, Gyeonggi-do 18448, Republic of Korea}

\author{Geonyoung Kim}
  \affiliation{Department of Electrical and Computer Engineering, Seoul National University, Seoul 08826, Republic of Korea}

\author{Kibog Park}
  \affiliation{Department of Physics, Ulsan National Institute of Science and Technology (UNIST), Ulsan 44919, Republic of Korea}
  \affiliation{Department of Electrical Engineering, Ulsan National Institute of Science and Technology (UNIST), Ulsan 44919, Republic of Korea}

\author{Yong-Ho Lee}
  \affiliation{Center for Superconducting Quantum Computing Systems, Korea Research Institute of Standards and Science, Daejeon 34113, Republic of Korea}

\author{Seungyong Hahn*}
  \affiliation{Department of Electrical and Computer Engineering, Seoul National University, Seoul 08826, Republic of Korea}
  \email[Authors to whom correspondence should be addressed: ]{ghchoi@kriss.re.kr}{hahnsy@snu.ac.kr}
\date{\today}

\begin{abstract}
Engineering the admittance of external environments connected to superconducting qubits is essential, as increasing the measurement speed introduces spontaneous emission loss to superconducting qubits, known as Purcell loss. 
Here, we report a {\crg broadband} Purcell filter within a small footprint, which effectively suppresses Purcell loss without losing the fast measurement speed.
We characterize the filter's frequency response at 4.3~K and also estimate Purcell loss suppression by finite-element-method simulations of superconducting planar circuit layouts with the proposed filter design.
{\crg The filter is fabricated with 200~nm-thick niobium films and shows} the measured bandwidth over 790~MHz within {\crg 0.29~mm$^2$ of compact size owing to densely packed spiral resonators.} 
The estimated lifetime enhancement {\crg indicates the effective protection of the qubit from Purcell loss.}
The presented filter design is expected to be easily integrated on existing superconducting quantum circuits for fast and multiplexed readout without occupying large footprint.
\end{abstract}

\maketitle
In circuit quantum electrodynamics~(cQED), Josephson junction based superconducting qubits are utilized as artificially engineered atoms~\cite{krantz2019quantum} and quantum non-demolition dispersive readout of qubit state is performed by probing the coupled readout resonator's frequency shift~\cite{krantz2019quantum,Koch2007charge, blais2004cavity, wallraff2005approaching}.
Recent progresses in scaling up the number of qubits in superconducting quantum circuits~\cite{Jurcevic2021demonstration, Kim2023evidence, Gold2021entanglement,gong2021quantum}, various quantum algorithms~\cite{Montanaro2016quantum}, and error correction protocols~\cite{Barends2014superconducting, google2023suppressing, krinner2022realizing, zhao2022realization} imply that the fast and high-fidelity multiplexed readout of qubits with distributed readout resonator frequencies is favorable to avoid complex signal wiring~\cite{chen2012multiplexed} and tight cryogenic cooling budgets~\cite{Krinner2019engineering}. 
In order to minimize crosstalk between the readout resonators, the resonators must be sufficiently detuned to each other, which necessitates a broad measurement bandwidth~\cite{heinsoo2018rapid}. 
However, increasing measurement speed to improve signal-to-noise ratio~(SNR) usually introduce {\crg larger} spontaneous emission loss in qubits, known as Purcell loss, leading to short qubit lifetime~$T_1$.

When a qubit is dispersively coupled to its readout resonator only, Purcell loss decay rate~$\Gamma_P=1/T_{1,P}$, an inverse of the Purcell limited lifetime~$T_{1,P}$, can be approximated as~\cite{Koch2007charge}:
\begin{equation}
  \label{eq:purcell_rate}
  \Gamma_P = \kappa_r\left(\frac{g_r}{\Delta_{qr}}\right)^2,
\end{equation}
where $\kappa_r$, $g_r$, and $\Delta_{qr}$ are readout resonator's decay rate, coupling strength, and frequency detuning between qubit and readout resonator, respectively.
From \hyperref[eq:purcell_rate]{Eq.~(1)}, the Purcell loss limited lifetime~$T_{1,P}$ decreases linearly with $\kappa_r$, while qubit state measurement speed exhibits nearly linear growth with $\kappa_r$~\cite{heinsoo2018rapid, jeffrey2014fast}.
If a qubit experiences various loss channels that exceed $\Gamma_P$, Purcell loss can be ignored by choosing small $\kappa_r$ to allow long $T_{1,P}$.
With the recent efforts to elucidate various loss channels in qubit, there have been remarkable improvements in qubit design~\cite{Pan2022engineering, Wang2022towards, eun2023shape} and fabrication process~\cite{murray2021material, Place2021newmaterial, osman2021simplified} that the importance of enigineering the Purcell loss is increasing to realize a practical quantum computer.
To protect qubits from the Purcell loss, Purcell filters have been developed to impede photon emission from qubits to the connected external circuits at qubit frequency while allowing it at the readout frequency.

One of the most commonly utilized Purcell filter types is the use of a single pole bandpass filter~\cite{jeffrey2014fast} or a bandstop filter coupled to readout resonators~\cite{reed2010fast}. 
Typical single-pole Purcell filters, however, have limited readout bandwidth that is not suitable for multiplexed readout scheme where readout resonators are sufficiently detuned~\cite{gambetta2006qubit}.
To overcome the narrow bandwidth of single-pole Purcell filter, coupling to individual single-pole Purcell filters engineered for every target readout resonator has {\crg shown effective cross-dephasing protection in a multiplexed readout scheme} but requires large footprint size~\cite{heinsoo2018rapid}. 
Meanwhile, the intrinsic single-pole Purcell filter has shown a remarkable Purcell loss suppression and a sufficient bandwidth but requires {\crg a careful engineering of the coupler position}~\cite{sunada2022fast}.
Another way to achieve a {\crg broadband} readout is to develop multi-pole Purcell filters such as stepped impedance bandpass filter~\cite{bronn2015broadband} and  mechanical ladder filter~\cite{cleland2019mechanical}. 
However, previously demonstrated multi-pole Purcell filters usually occupy a large footprint~\cite{bronn2015broadband} or requires additional fabrication process~\cite{cleland2019mechanical} where on-chip integration may become cumbersome. 
Therfore, for the qubit lifetime to reach the intrinsic limit of the qubit, a Purcell filter supporting broadband bandwidth, compact-size, and compatible to the standard superconducting qubit circuit fabrication process is necessary.

Here, we introduce a 4-pole Purcell filter that has been designed following coupled resonator filter synthesis techniques from microwave engineering~\cite{cameron2018microwave}. 
The filter comprises of symmetric, densely packed superconducting spiral coplanar-waveguide~(CPW) resonators~\cite{kanaya2001miniatureized, ma2006miniaturized} coupled to the neighboring spiral CPW resonators. 
We highlight the significant features of the spiral CPW Purcell filter design including its compact footprint of 0.29~mm$^2$, comparable to a typical transmon qubit footprint size~\cite{Wang2022towards,heinsoo2018rapid}, wide bandwidth of 790 MHz centered at 6.93 GHz in the passband for multiplexed readout application, and large attenuation at frequency out of the passband for suppressed Purcell loss and the fast readout application. 

\begin{figure}[!t]
  \includegraphics[clip,width=0.99\columnwidth]{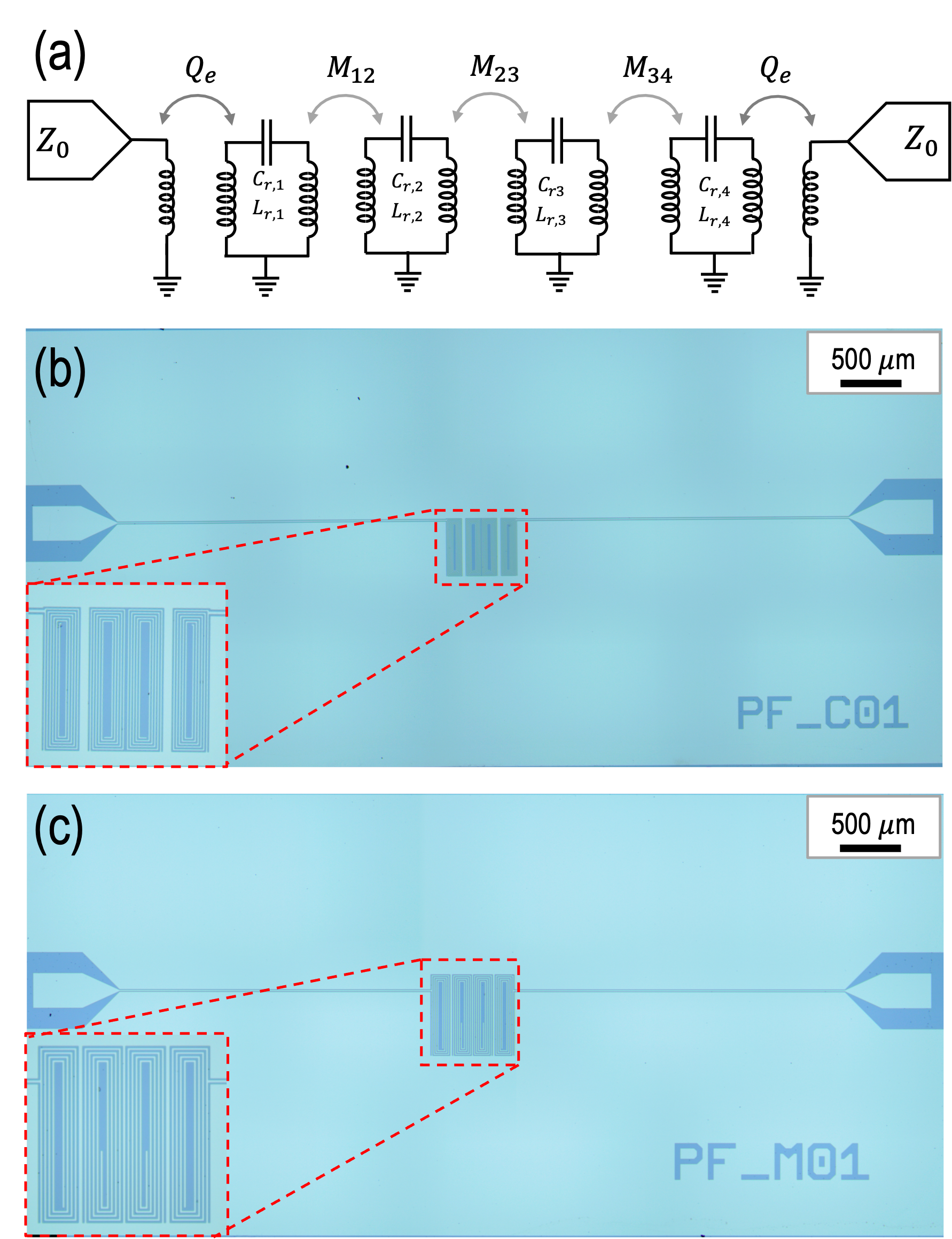}
  \caption{(a) A schematic of 4 {\crg direct inductively} coupled resonator band-pass filter with characteristic impedance $Z_0$ of ports. (b) and (c) are the optical microscope images of PF-C and PF-M chip respectively. Dark blue regions represent the etched patterns while light blue regions represent the unetched niobium films. The inset figures in (b) and (c) show the enlarged Purcell filter geometries of $500~\mu\textrm{m}\times580~\mu\textrm{m}$ and $680~\mu\textrm{m}\times700~\mu\textrm{m}$ respectively.}
  \label{fig:1} 
\end{figure}

\hyperref[fig:1]{Figure~1(a)} shows circuit diagram of the 4-pole bandpass filter using the Chebyshev polynomials~\cite{pozar2000microwave, cameron2018microwave} connected to the port characteristic impedance~$Z_0=50~\Omega$. 
To minimize footprint, the coupled resonator filter synthesis techniques are chosen~\cite{cameron2018microwave, ma2006miniaturized}.
Here, $C_{r,i}$ and $L_{r,i}$ are the capacitance and inductance of the $i^\textrm{th}$ coupled CPW resonator, respectively.
{\crg Mutual inductive} coupling coefficient between resonator $i$ and $j$ are denoted as $M_{ij}$ while coupling to the input and output ports are {\crg equivalently} defined as an external quality factor~$Q_e$.
Design parameters of $M_{ij}$ and $Q_e$ can be calculated according to the desired targets (1) center frequency~$\omega_c/2\pi$; (2) 3~dB bandwidth~$B/2\pi$, where the transmission magnitude reduces by 3~dB; (3) the maximum ripple magnitude of the passband.
We design two Purcell filters: (1) PF-C with $\omega_c/2\pi=7.05$~GHz and $B/2\pi=850$~MHz design targets; (2) PF-M with $\omega_c/2\pi=6.91$~GHz and $B/2\pi=970$~MHz design targets to be integrated into typical multiplexed readout schemes.
{\crg PF-C is designed to have higher $\omega_c/2\pi$ by choosing smaller inductance compared to PF-M.}
Both filters are designed to have 0.5~dB of the maximum ripple in the passband.
Filters are designed with the help of Qiskit-Metal~\cite{Qiskit_Metal}, an open source Python package for superconducting quantum circuit design, while simulations are performed using Ansys HFSS, {\crg3D electromagnetic finite-element-method~(FEM) software}~\cite{Ansys}.
We assume that the relative permittivity~$\varepsilon_r$ of silicon substrate is 11.9 when designing the filters.

\begin{figure}[!t]
  \includegraphics[clip,width=0.99\columnwidth]{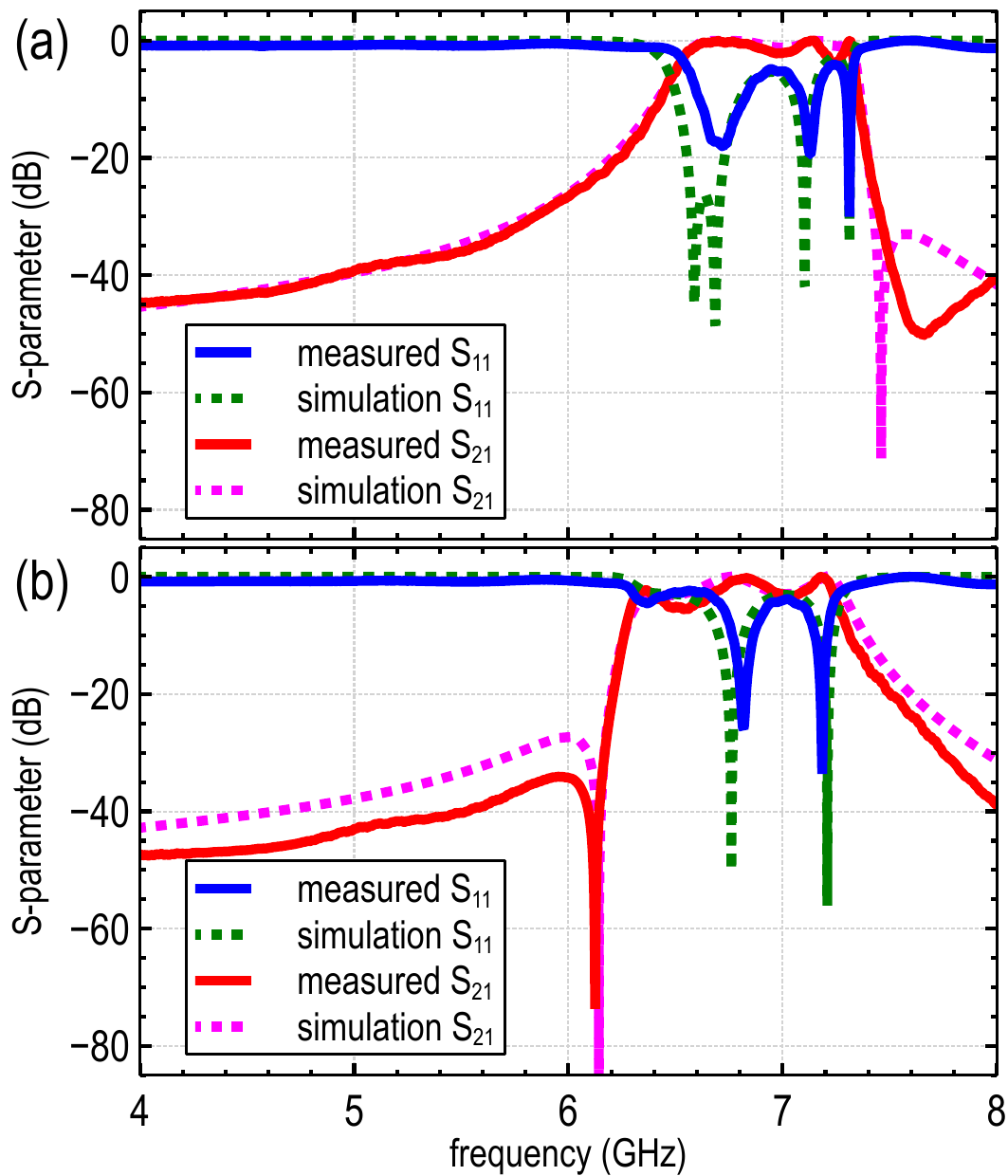}
  \caption{Measured scattering parameters of (a) PF-C and (b) PF-M chips at 4.3~K. Measured $\textrm{S}_{11}$ and $\textrm{S}_{21}$ values are plotted with solid blue and red lines respectively while simulation $\textrm{S}_{11}$ and $\textrm{S}_{21}$ are plotted with dotted {\crg green and magenta lines respectively. The simulation are conducted in consideration of kinetic inductance from niobium film layers and relative permittivity, $\varepsilon_r=11.45$, of silicon substrate at low temperatures.}}
  \label{fig:2} 
\end{figure}

The 4-pole Purcell filters are fabricated using DC magnetron sputtered and dry etched 200~nm-thick superconducting niobium film on a 375~$\mu$m-thick and 3 inch-diameter high resistive silicon wafer.
The chips are designed to be $4~\textrm{mm}\times 8~\textrm{mm}$ in overall size to fit in the microwave package, regardless of the Purcell filter's footprint size.
The fabricated Purcell filter sample chips are shown in \hyperref[fig:1]{Fig.~1(b)} and \hyperref[fig:1]{1(c)}.
To characterize the scattering parameters~($\textrm{S}_{11}$, $\textrm{S}_{21}$) of the Purcell filters, PF-C and PF-M chips are mounted on the copper microwave package and wire-bonded with aluminum wires {\crg to suppress the unwanted slot-line modes~\cite{chen2014fabrication,Endo2016}}.
We measure the filters' scattering parameters with calibrated Keysight E5071C vector network analyzer~(VNA) and perform the short-open-load-transmission calibration with Keysight 85521A calibration kit before the measurements at liquid-helium boiling temperatures of 4.3~K.
Also, transmission through the measurement setup system is subtracted from the output of the calibrated VNA~\cite{sage2015study}.
The measured and simulated scattering parameters of PF-C and PF-M are shown in \hyperref[fig:2]{Fig.~2(a)} and \hyperref[fig:2]{2(b)}, respectively.
To achieve fast and high-fidelity multiplexed qubit measurements, with a target $\kappa_r/2\pi$ of 10~MHz and a target spacing between readout resonator frequencies to be 10 times larger than $\kappa_r/2\pi$~\cite{heinsoo2018rapid,Chen2023transmon}, the PF-C filter can be integrated into circuits with 7 readout resonators, while the PF-M filter can be integrated into circuits with 9 readout resonators.

The measured Purcell filters' transmissions show increased ripples in the passband while decreased $\omega_c$ and $B$ compared to design targets.
Possible contributions are due to the impedance mismatch in the microwave package~\cite{huang2021microwave}, changes in $\epsilon_r$ of silicon substrate at low temperatures~\cite{li2023experimentally}, or kinetic inductance from the thin superconducting film~\cite{Niepce2020geometric}.
Moreover, the kinetic inductance originating from the thin superconducting niobium film layer, becomes non-negligible in our compact and narrow-gap Purcell filter designs due to the effectively increasing geometric length and decreasing resonant frequency~\cite{nigg2012blackbox}.
In order to investigate the possible contributions, we model a silicon substrate with $\varepsilon_r=11.45$~\cite{krupka2006measurements} and apply impedance boundary conditions, where the surface impedance is non-zero for superconducting layers.
Calculation of superconducting niobium film's complex conductivity is done with Mattis-Bardeen equations~\cite{linden1994modified, park2021design, gao2008physics}.
The effective London penetration depth~$\lambda_\textrm{eff}$ is calculated to be 142~nm at $T=4.3$~K, which can be described as a complex permittivity~$\epsilon_r$ of superconducting layers~\cite{Niepce2020geometric, li2023experimentally}.
We find that our niobium film is in the local limit~\cite{gao2008physics} due to large $\lambda_\textrm{eff}$ compared to the previous studies~\cite{henkels1977penetration, gubin2005dependence, Niepce2020geometric}.
Our simulation results {\crg with kinetic inductance from niobium film layers and low relative permittivity of the substrate} show good agreements with the measured $\textrm{S}_{11}$ and $\textrm{S}_{21}$ results as shown in \hyperref[fig:2]{Fig.~2}. 
Specifications of PF-C and PF-M are summarized in \hyperref[table:1]{Table.~1}.

\begin{table}[!t]
\caption{\label{table:1} Measurement, and simulation results of characteristic parameters of PF-C and PF-M filters. {\crg Simulations are performed in consideration of kinetic inductance from superconducting film and dielectric constant of silicon substrate at 4.3~K}.}
\begin{ruledtabular}
  \begin{tabular}{llll}
  & Parameters & PF-C & PF-M \\ \hline  
  \multirow{2}{*}{measurement} & $\omega_c/2\pi$& 6.93~GHz & 6.78~GHz\\
  & $B/2\pi$ & 794~MHz & 915~MHz\\
  \multirow{2}{*}{simulation} & $\omega_c/2\pi$ & 6.94~GHz & 6.82~GHz\\
  & $B/2\pi$ & 794~MHz& 954~MHz 
  \end{tabular}
\end{ruledtabular}
\end{table}

The decay rate to the external environments connected to a superconducting qubit can be calculated {\crg from the} real part of admittance~{\crg$Y_\textrm{ext}(\omega_q)$ seen by} the Josephson junction of the qubit at frequency~$\omega_q$.
For a transmon qubit with a fixed $E_c$, Purcell limited lifetime~$T_{1,P}$ of the qubit in an arbitrary circuit can be expressed as~\cite{bronn2015broadband,cleland2019mechanical,sunada2022fast}:
\begin{equation}
  \label{eq:purcell_lifetime}
  T_{1,P} = \frac{C_\Sigma}{\textrm{Re}\left[{\crg Y_\textrm{ext}}(\omega_q)\right]},
\end{equation}
where qubit capacitance {\crg is} $C_\Sigma=e^2/2E_c$ and $e$ is the {\crg charge of an electron}.
We substitute {\crg the} Josephson junction with a lumped port to calculate the real part of {\crg $Y_\textrm{ext}$} in a wide frequency range. 
In {\crg Ansys} HFSS simulation, lossless network is assumed to calculate the Purcell loss only.
\begin{figure}[!t]
  \includegraphics[clip,width=0.97\columnwidth]{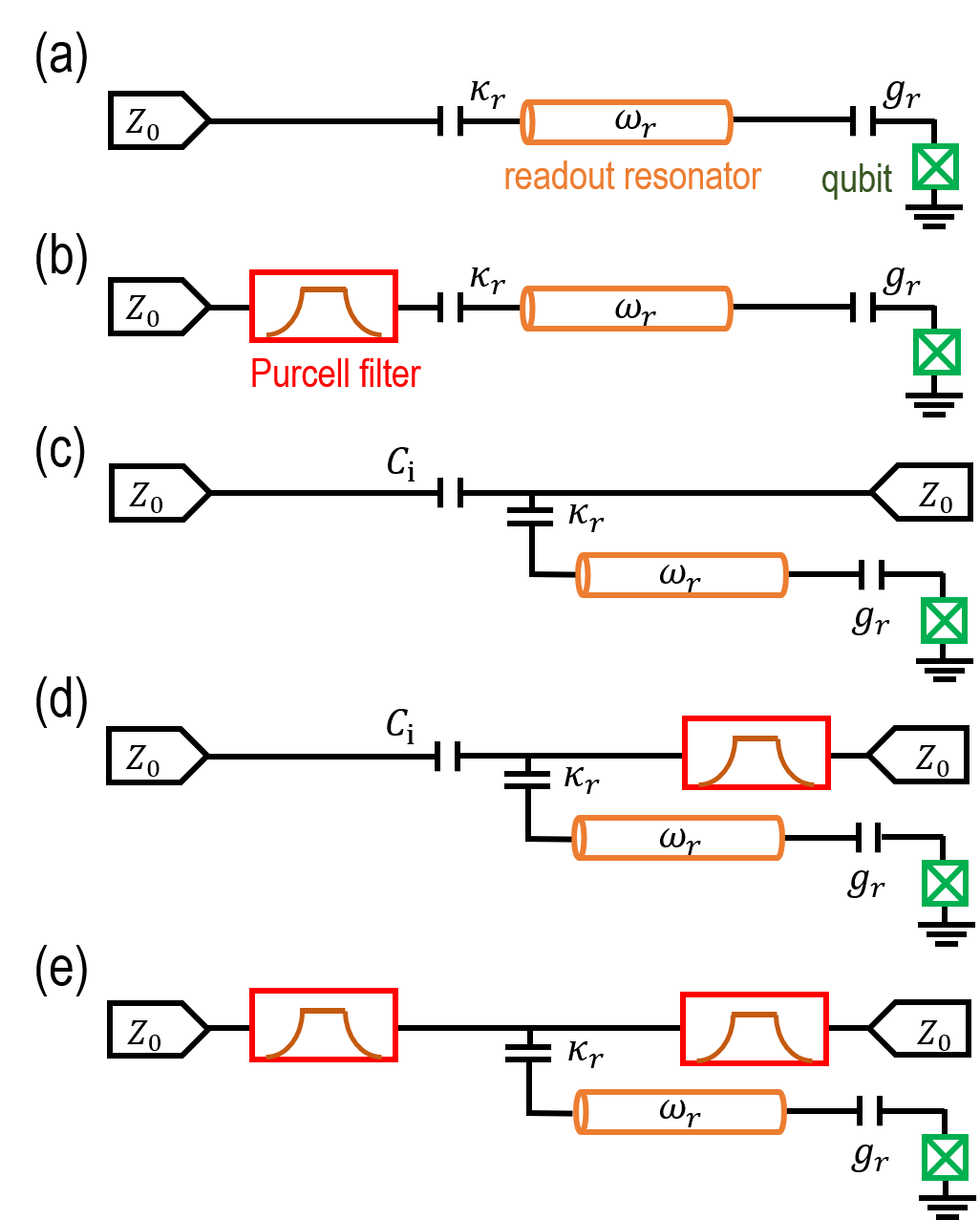}
  \caption{Circuit schematics of (a) single-port readout without Purcell filter, (b) single-port readout with Purcell filter, (c) two-port readout without Purcell filter, (d) two-port readout with Purcell filter inserted at output port, and (e) two-port readout with Purcell filters inserted at input and output ports. 
  Purcell filters, readout resonators, and qubits are illustrated with red, orange, green layers respectively.}
  \label{fig:3} 
\end{figure}
{\crg To quantitatively evaluate the Purcell loss suppression in qubits, the figure of merit~(FOM) at detuning frequency between a qubit and a readout resonator, $\Delta_{qr}=\omega_q-\omega_r$, is defined as:
\begin{equation}
  \textrm{FOM}(\Delta_{qr}) = \frac{T_{1,P}~\textrm{with Purcell filter}}{T_{1,P}~\textrm{without Purcell filter}},
\end{equation}
where $T_{1,P}$ with (or without)} Purcell filter is calculated by {\crg Ansys HFSS} simulations on geometry where a superconducting qubit, capacitively coupled to a readout CPW resonator, is lying on the single-port readout circuit or two-port readout circuit as seen in \hyperref[fig:3]{Fig. 3}.

To calculate the FOM for Purcell loss suppression, single~(or two-port readout) circuits with~(or without) Purcell filter are simulated to verify the filter's performance in various circuit configurations.
The Purcell filter~(PF-C) is merged into the bare single qubit readout circuits of \hyperref[fig:3]{Fig.~3(a)} and \hyperref[fig:3]{3(c)}. 
For two-port readout circuits, an input capacitor~$C_\text{i}=30$~fF is inserted at the input port of the readout transmission line for the directionality to the output port by suppressing the reflected signal flowing into the input port~\cite{heinsoo2018rapid}.
Moreover, a two-port readout circuit, where Purcell filters are inserted at the input and output ports, is also studied to maximize the FOM as shown in \hyperref[fig:3]{Fig.~3(e)}. 
The CPW readout resonator geometry is adjusted to have resonant frequency,~$\omega_r/2\pi\approx7$~GHz, which is near the center of our Purcell filter's passband for single-port readout circuits and two-port readout circuits.
Next, charging energy~$E_c/2\pi$ of floating transmon qubit~\cite{sete2021floating} and coupling strength to the readout resonator~$g_r/2\pi$ is designed to reflect practical values for typical superconducting qubit experiments while $\kappa_r/2\pi$ is designed to be large enough for fast qubit state measurement. 
$E_c$ and $g_r$ are calculated by extracting the capacitance matrix of the circuit layout.
Parameter values for the circuits can be found in \hyperref[fig:4]{Fig.~4} while the images of the simulation circuit layouts can be found in supplementary material.
We illustrate the simulated Purcell loss limited lifetime~$T_{1,P}$ of various circuits (as shown in \hyperref[fig:3]{Fig.~3}) with the readout bandwidths of the integrated Purcell filters highlighted in cyan background color in \hyperref[fig:4]{Fig.~4}.

\begin{figure}[!t]
  \includegraphics[clip,width=0.99\columnwidth]{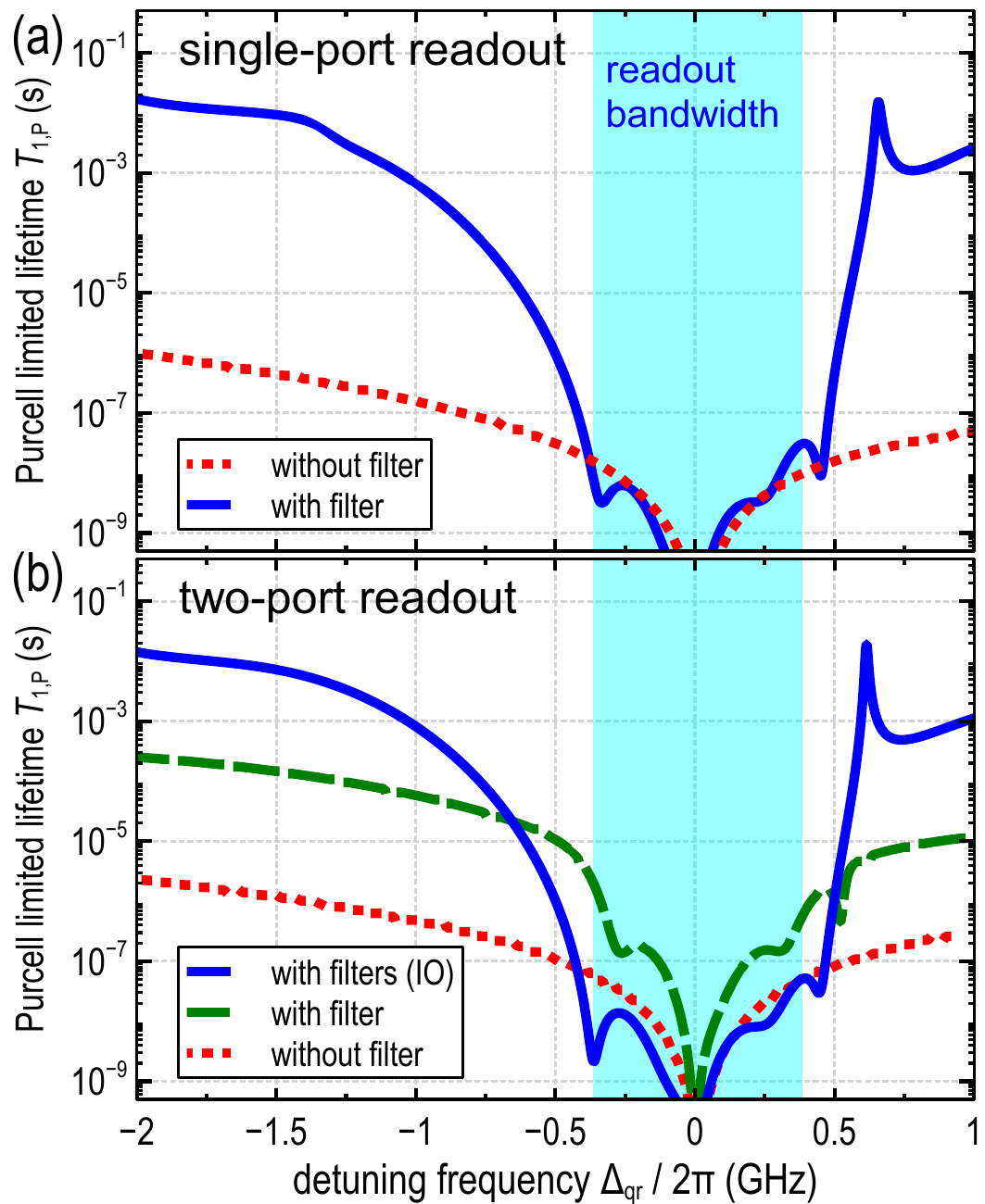}
  \caption{Simulation results of Purcell loss limited lifetime~$T_{1,P}$ for (a) $T_{1,P}$ of single-port readout circuit without Purcell filter~(dotted light blue) and with Purcell filter~(solid dark blue). Here, readout resonator frequency is $\omega_r/2\pi=7.08$~GHz while $E_c/2\pi$, $g_r/2\pi$, and $\kappa_r/2\pi$ are designed to have 258~MHz, 175~MHz, and 30.8~MHz, respectively. (b) $T_{1,P}$ of two-port readout circuit without Purcell filter~(dotted light blue), with a Purcell filter at the output port~(dash dark blue), and with Purcell filters at the input and output~(IO) ports~(solid blue). Here, readout resonator frequency is $\omega_r/2\pi=7.05$~GHz while $E_c/2\pi$, $g_r/2\pi$, and $\kappa_r/2\pi$ are designed to have 263~MHz, 123~MHz, and 25.1~MHz, respectively.}
  \label{fig:4} 
\end{figure}

Simulated $T_{1,P}$ for the single-port readout circuits in \hyperref[fig:3]{Fig.~3(a)-(b)} are plotted as a function of detuning frequency,~$\Delta_{qr}$, between a qubit and a readout resonator in \hyperref[fig:4]{Fig.~4(a)}.
Within a readout bandwidth of the filter, Purcell loss is not suppressed well compared to the bare limit of $T_{1,P}$ as the filter allows photon loss from the qubit into the transmission line through the readout resonator.
Small dips within the readout bandwidth are due to the coupling between the readout resonator and spiral CPW resonator in Purcell filter.
It is notable that the {\crg Purcell loss can be suppressed as the absolute $|\Delta_{qr}/2\pi|$ increases since, the Purcell filter demonstrates a sharp $\textrm{S}_{21}$ suppression just before entering the readout bandwidth.}
Thus, for a single-port readout circuit application, it's advantageous to engineer $|\Delta_{qr}/2\pi|$ to be equal to or greater than {\crg 0.5}~GHz to effectively suppress Purcell loss in a superconducting qubit.
In particular, for {\crg the $|\Delta_{qr}/2\pi|\approx1~$GHz, $T_{1,P}\approx670~\mu$s} can be achieved.
{\crg This performance can be compared to that of a typical single-pole bandpass Purcell filter~\cite{jeffrey2014fast}, where the quality-factor of the quarter-wavelength CPW filter is $Q_F=30$ and the expected $T_{1,P}=14.5~\mu$s for the same values of $\omega_r, \kappa_r, g_r,$ and $E_c$.}

Simulated $T_{1,P}$ for the two-port readout circuits in \hyperref[fig:3]{Fig.~3(c)-(e)} are plotted as a function of detuning frequency~$\Delta_{qr}$ in \hyperref[fig:4]{Fig.~4(b)}. 
{\crg The} two-port readout circuit with a Purcell filter at the output port shows rapid Purcell loss suppression that FOM{\crg($\Delta_{qr}/2\pi=-1$~GHz)} over 89 is achieved but the decreased rate of change for Purcell loss suppression is observed as $|\Delta_{qr}|$ increases.
The two-port readout circuit with a single Purcell filter experiences $T_{1,P}$ dip at $\Delta_{qr}/2\pi\approx0.5$~GHz.
The dip is found to be attributed to the resonance of long transmission line between spiral CPW resonator in filter and capacitor at input port.
We find that if transmission line length between the Purcell filters varies, the dip shows different resonant frequency that engineering the unwanted Purcell loss can be removed.
Other dips are attributed to the Purcell filter's wide passband.
In \hyperref[fig:4]{Fig.~4(b)}, simulated $T_{1,P}$ for the two-port circuit with two Purcell filters at IO ports is also presented.
It is notable that two-port circuit with two Purcell filters at IO ports shows nearly quadratic increments of $T_{1,P}$ outside of the readout bandwidth that FOM over 1500 can be continuously achieved for $|\Delta_{qr}/2\pi|>1$~GHz.
In particular, $T_{1,P}$ can be improved up to 13.9~ms~($\textrm{FOM}({\crg \Delta_{qr}/2\pi=-2~\textrm{GHz}})=5265$) which is far above the current state-of-the-art qubit lifetime~\cite{Wang2022towards,somoroff2023millisecond}, whereas bare qubit lifetime is only 2.34~$\mu\textrm{s}$ due to the large $\kappa_r$ and $g_r$. 
{\crg If a qubit is limited by the surface dielectric loss~\cite{jeffrey2014fast, murray2021material} and has $T_1\approx 100~\mu$s, choosing smaller spiral CPW resonator geometry and less pole number will be enough to suppress Purcell loss with $T_{1,P}\approx 1~$ms target design.
}

In conclusion, we have proposed and characterized broadband Purcell filter designs with compact footprint size for fast and multiplexed superconducting qubit measurements.
We have fabricated Purcell filters which are measured to have bandwidth over than 790~MHz while occupying 0.29~$\textrm{mm}^2$ footprint size. 
Our design also allows alternative implementation of filters with a broader bandwidth or smaller footprint size by choosing smaller gap and linewidth of spiral CPW resonators. 
{\crg The simulation results of the filter scattering parameters show that a reduction in substrate permittivity and the presence of kinetic inductance from superconducting layers can contribute to alterations in both the center frequency and bandwidth of the passband.} 
Thus, discrepancies between design and fabricated device can be further minimized by considering accurate material properties at low temperatures.
Simulations on superconducting quantum circuits have affirmed that the filter effectively improves the Purcell limited lifetime $T_{1,P}$ well above the reported state-of-the-art qubit lifetimes.
Our simulation results also have shown that the Purcell loss can be reduced {\crg in two-port readout chips} by integrating multiple Purcell filters.
This compact, broadband Purcell filter design is expected to facilitate fast, high-fidelity, and multiplexed qubit state {\crg readout} by seamlessly {\crg integrating} into superconducting quantum circuits without increasing the chip layout size {\crg or} requiring additional fabrication process.

{\crg 
See the supplementary material for additional details on the superconducting circuit simulation methods with the Purcell filter to extract Purcell loss of the qubit as well as the detailed chip layout geometries.
}

This research was supported in part by Samsung Electronics.
This research was also supported by National R\&D Program through the National Research Foundation of Korea~(NRF) funded by Ministry of Science and ICT~(2022M3I9A1072846) and in part by the Applied Superconductivity Center, Electric Power Research Institute of Seoul National University. 
This research was supported by Development of quantum-based measurement technologies funded by Korea Research Institute of Standards and Science~(KRISS–2023–GP2023-0013).

\bibliography{shp_reference}
\end{document}